\documentclass[runningheads]{llncs}
\usepackage{graphicx}
\usepackage{listings}
\usepackage{xcolor}
\usepackage{subfig}
\usepackage[square,numbers]{natbib}
\usepackage{multirow}
\usepackage{tabularx}
\usepackage{comment}
\usepackage{amsmath}
\usepackage[labelfont=bf, justification=justified]{caption}
\usepackage[htt]{hyphenat}
\definecolor{codegreen}{rgb}{0,0.6,0}
\definecolor{codegray}{rgb}{0.5,0.5,0.5}
\definecolor{codepurple}{rgb}{0.58,0,0.82}
\definecolor{backcolour}{rgb}{0.95,0.95,0.92}

\lstdefinestyle{mystyle}{
    backgroundcolor=\color{backcolour},   
    commentstyle=\color{codegreen},
    keywordstyle=\color{magenta},
    numberstyle=\tiny\color{codegray},
    stringstyle=\color{codepurple},
    basicstyle=\ttfamily\scriptsize,
    breakatwhitespace=false,         
    breaklines=true,                 
    captionpos=b,                    
    keepspaces=true,                 
    numbers=left,                    
    numbersep=5pt,                  
    showspaces=false,                
    showstringspaces=false,
    showtabs=false,                  
    tabsize=2
}
\newcommand{\onemkl}{\text{oneMKL}}
\newcommand{\curand}{\text{cuRAND}}
\newcommand{\hiprand}{\text{hipRAND}}

\begin{document}
\title{Achieving near native runtime performance and cross-platform performance portability for random number generation through SYCL interoperability\thanks{With support from the DOE HEP Center for Computational Excellence at Lawrence Berkeley National Laboratory under B\&R KA2401045.}}
\titlerunning{Performance portable RNG with SYCL interoperability}
%
\author{Vincent R. Pascuzzi\inst{1}\thanks{Now at Brookhaven National Laboratory, Upton, NY 11973 USA.}\orcidID{0000-0003-3167-8773} \and \\
Mehdi Goli\inst{2}\orcidID{0000-0002-2774-0821}}
\authorrunning{V.R. Pascuzzi and M. Goli}
%
\institute{Lawrence Berkeley National Laboratory, Berkeley CA 94590, USA
\email{pascuzzi@bnl.gov} \and
Codeplay Software Ltd., Edinburgh EH3 9DR, UK \\
\email{mehdi.goli@codeplay.com}}
\maketitle              
\begin{abstract}
High-performance computing (HPC) is a major driver accelerating scientific research and discovery, from quantum simulations to medical therapeutics.
While the increasing availability of HPC resources is in many cases pivotal to successful science, even the largest collaborations lack the computational expertise required for maximal exploitation of current hardware capabilities.
The need to maintain multiple platform-specific codebases further complicates matters, potentially adding constraints on machines that can be utilized.
Fortunately, numerous programming models are under development that aim to facilitate portable codes for heterogeneous computing.
One in particular is SYCL, an open standard, C++-based single-source programming paradigm.
Among SYCL's features is interoperability, a mechanism through which applications and third-party libraries coordinate sharing data and execute collaboratively.
In this paper, we leverage the SYCL programming model to demonstrate cross-platform performance portability across heterogeneous resources.
We detail our NVIDIA and AMD random number generator extensions to the oneMKL open-source interfaces library.
Performance portability is measured relative to platform-specific baseline applications executed on four major hardware platforms using two different compilers supporting SYCL.
The utility of our extensions are exemplified in a real-world setting via a high-energy physics simulation application.
We show the performance of implementations that capitalize on SYCL interoperability are at par with native implementations, attesting to the cross-platform performance portability of a SYCL-based approach to scientific codes.

\end{abstract}
\section{Introduction}
The proliferation of heterogeneous systems in high-performance computing (HPC) is providing scientists and researchers opportunities to solve some of the
world's most important and complex problems.
Coalescing central processing units (CPU), co-processors, graphics processing units (GPU) and other hardware accelerators with high-throughput inter-node networking capabilities has driven science and artificial intelligence through insurmountable computational power.
Industry continues to innovate in the design and development of
increasingly performant architectures and platforms, with each vendor typically commercializing a myriad of proprietary libraries optimized for their
specific hardware.
What this means for physicists and other domain scientists is that codes need to be translated, or ported, to multiple languages, or adapted to some specific programming model for best performance.
While this could be a useful and instructive exercise for some, many are often burdened by their limited numbers of developers that can write such codes.
Fortunately, as a result of the numerous architectures and platforms,
collaborative groups within academia, national laboratories and even industry
are developing portability layers atop common languages that aim to target a variety of vendor hardware.
Such examples include Kokkos~\cite{edwards2014kokkos} (Sandia National Laboratory, USA), RAJA~\cite{hornung2014raja} (Lawrence Livermore National Laboratory, USA) and SYCL~\cite{sycl2020} (Khronos Group).

Mathematical libraries are crucial to the development of scientific codes.
For instance, the use of random numbers in scientific applications, in particular high-energy physicists (HEP) software, is almost
ubiquitous~\cite{james2020review}.
For example, HEP experiments typically have a number of steps that are required as part of their Monte Carlo (MC) production: event generation, simulation, digitization and reconstruction.
In the first step, an MC event generator~\cite{Buckley_2011} produces the outgoing particles and their four-vectors given some physical process.
Here, random numbers are used for, \textit{e.g.}, sampling initial state kinematics and evaluating cross sections.
Simulation software, \textit{e.g.}, Geant4~\cite{Agostinelli:2002hh} and FastCaloSim~\cite{Schaarschmidt2017,10.3389/fdata.2021.665783} from the ATLAS Experiment~\cite{Aad:1129811}, require large quantities of random numbers for sampling particle energies and secondary production kinematics, and digitization requires detector readout emulation, among others.
With the rise of machine learning, random number production is required even at the analysis level~\cite{feickert2021living}.
\subsection{Contribution}
The focus of this paper is to evaluate the cross-platform performance portability of SYCL's interoperability functionality using various closed-source vendor random number generation APIs within a single library, and analyze the
performance of our implementation in both artificial and real-world applications.

To achieve this, we have:

\begin{itemize}
\item developed SYCL-based random number generator (RNG) implementations within the \onemkl{} open-source interfaces library that can target both AMD and NVIDIA GPUs, two major HPC hardware providers;
\item evaluated the performance portability of the proposed solution for RNG on  Intel and AMD CPUs, and Intel, AMD and NVIDIA GPUs to investigate the performance overhead of the abstraction layer introduced by the SYCL API;
\item integrated our RNG implementations into FastCaloSim to further investigate the applicability of the proposed solution on an existing real-world application for high-energy physics calorimeter simulations, currently relying on various vendor-dependent libraries. The RNG routine for the original version of FastCaloSim can only run on x86 CPU architectures and NVIDIA GPUs using \curand{}. The integration proposed here enabled on-device generation of random numbers per FastCaloSim simulation event to additionally target all current HPC vendor hardware from a single entry point; and
\item analyzed the cross-platform performance portability by comparing the SYCL-based implementation of FastCaloSim to the original C++-based and CUDA codes which use native vendor-dependent RNG to investigate possible performance overheads associated with SYCL interoperability.
\end{itemize}

Our work utilizes Data Parallel C++ (DPC++)~\cite{dpcpp} and hipSYCL~\cite{alpay2020sycl}, two different existing LLVM-based SYCL compilers, capable of providing plug-in interfaces for CUDA and HIP support as part of SYCL 2020 features that enable developers to target NVIDIA and AMD GPUs, respectively.

The rest of this paper is organized as follows.
Section~\ref{sec:rel-work} discusses existing parallel frameworks and libraries providing functionalities used in scientific applications, along with our proposed solution to target the cross-platform portability issue. Section~\ref{sec:prelims} briefly introduces the SYCL programming model used in this work.
In Section~\ref{sec:support-curand}, we discuss more technical aspects and differences between the \curand{} and \hiprand{} APIs, and also details the implementation. Benchmarks for performance portability are described in Section~\ref{sec:perfport}, and our results are presented in Section~\ref{sec:results}. Lastly, Section~\ref{sec:conclusion} summarizes our work, and suggests potential extensions and improvements for future work.

\section{Related Work}
\label{sec:rel-work}
\subsection{Parallel Programming Frameworks}
Parallelism across a variety of hardware can be provided through a number different parallel frameworks, each having a different approach and programming style.
Typically written in C or C++, each framework provides different variations on the language, allowing programmers to specify the task parallel patterns.

Introduced by Intel, Thread Building Blocks (TBB)~\cite{pheatt2008intel} provides a C++-based template library supporting parallel programming on multi-core processors. TBB only support parallelism on CPUs, hence, parallel applications dependent on
TBB cannot be directly ported to GPUs or any other accelerator-based platform.

NVIDIA's CUDA~\cite{cuda} API is a C/C++-based low-level parallel programming framework exclusively for NVIDIA GPUs.
Its support of C++-based template meta programming features enables CUDA to provide performance portability across various NVIDIA devices, however, its lack of portability can be a barrier for research groups with access to other vendor
hardware.

OpenCL~\cite{stone2010opencl}, from the Khronos Group, is an open-standard cross-platform framework supported by various vendors and hardware platforms However, its low-level C-based interface could hinder the development of performance portability on various hardware.
Also from the Khronos Group is SYCL~\cite{sycl2020}, an open-standard C++-based programming model that facilitates the parallel programming on heterogeneous platforms. SYCL provides a single-source abstraction layer enabling developers to write both host-side and kernel code in the same file.
Employing C++-based template programming, developers can leverage higher level programming features when writing accelerator-enabled applications, having
the ability to integrate the native acceleration API, when needed, by using different interoperability interfaces provide by SYCL.

The Kokkos~\cite{edwards2014kokkos} and RAJA~\cite{hornung2014raja} abstraction layers expose a set of C++-based parallel patterns to facilitate operations such as parallel loop execution, reorder, aggregation, tiling, loop partitioning and kernel transformation. They provide C++-based portable APIs for users to alleviate the difficulty of writing specialized code for each system. The APIs can be mapped onto a specific backend -- including OpenMP, CUDA, and more recently SYCL -- at runtime to provide portability across various architectures.

\subsection{Linear Algebra Libraries}
There are several vendor-specific libraries which provide highly optimized linear algebra routines for specific hardware platforms.
The ARM Compute Library~\cite{arm} provides a set of optimized functions for linear algebra and machine learning optimized for ARM devices.
Intel provides MKL~\cite{mkl} for its linear algebra subroutines for accelerating BLAS, LAPACK and RNG routines targeting Intel chips, and
NVIDIA provides a wide ecosystem of closed source libraries for linear algebra operations, including cuBLAS~\cite{cublas} for BLAS routines, \curand{}~\cite{curand} for RNG and cuSPARSE~\cite{cusparse} for sparse linear algebra.
AMD offers a set of hipBLAS~\cite{hipblas} and \hiprand{}~\cite{hiprand} libraries atop the ROCm platform, which provide highly-optimized linear algebra routines for AMD GPUs.
Each of these libraries is optimized specifically for particular hardware architectures, and therefore do not provide portability across vendor hardware.

\onemkl{}~\cite{onemkl} is an community-driven open-source interface library developed using the SYCL programming model, providing linear algebra functionalities used in various domains such as high-performance computing, artificial intelligence and other scientific domains.
The front-end SYCL-based interface could be mapped to the vendor-optimized backend implementations either via direct SYCL kernel implementations or SYCL interoperability using built-in vendor libraries to target various hardware backends.
Currently, \onemkl{} supports BLAS interfaces with vendor-optimized backend implementations for Intel GPU and CPU, CUDA GPUs and RNG interfaces which wrap the optimized Intel routines targeting x86 architectures and Intel GPUs.

\subsection{The Proposed Approach}
There are numerous highly-optimized libraries implemented for different device-specific parallel frameworks on different hardware architectures and platforms. Several parallel frameworks provide parallel models which hide the memory hierarchies and execution policies on different hardware. This can be due to a lack of a common language to abstract away the memory and execution models from various heterogeneous devices, hence, leaving cross-platform performance portability of high-level applications a challenging issue and an active area of research.
Recent work in adopting SYCL~\cite{10.1145/3388333.3388643,costanzo2021early,10.1145/3318170.3318190} as the unifying programming model across different hardware can be considered as a viable approach to develop a cross-platform performance portable solution targeting various hardware architectures while sharing the same interface.
More specifically, SYCL interoperability with built-in kernels enables vendors to use a common unifying interface, to ``glue-in'' their optimized hardware-specific libraries for current and next generations of processors and accelerators.

In this paper, we leverage the SYCL programming model and interoperability to enable cross-platform performance portable random number generator targeting major HPC hardware, including NVIDIA and AMD GPUs. The proposed solution has been integrated into the \onemkl{} open-source interfaces library as additional backends targeting these vendors, extending the library's portability and offering nearly native performance.
The applicability of the proposed approach was further studied in a high-energy physics calorimeter simulation software to evaluate the performance of the proposed abstraction method on a real-world scientific application.

\section{SYCL Overview}
\label{sec:prelims}
SYCL is an open-standard C++-based programming model that facilitates parallel programming on heterogeneous platforms. 
It provides a single source programming model, enabling developers to write both host-side and kernel code in the same file.
Employing C++-based template programming, developers can leverage higher-level programming features in writing accelerator-enabled applications with the ability to integrate the native acceleration API, when needed, by using different interoperability interfaces provided by SYCL.

A SYCL application is structured in three scopes that controls the flow, as well as the construction and lifetimes of the various objects used within it.
\begin{itemize} 

\item{\it Application scope}: specifies all other code outside of a command group scope.

\item{\it Command group scope}: specifies a unit of work that is comprised of a kernel function and accessors.

\item{\it  Kernel scope}: specifies a single kernel function to interface with native objects and is executed on the device. 
\end{itemize}
To execute a SYCL kernel on an accelerator device, \textit{command groups} containing the kernel must be submitted to a SYCL queue. 
When a command group is submitted to a queue, the SYCL runtime system tracks data dependencies and creates (expands) a new (existing) dependency graph --
a directed acyclic graph (\textit{DAG}) -- to orchestrate kernel executions.
Once the dependency graph is created, the correct ordering of kernel execution on any available device is guaranteed by the
SYCL runtime system via a set of rules defined for dependency checking.

Interoperability is enabled via the aforementioned low-level APIs by facilitating the SYCL runtime system's interaction with native objects for the supported backends~\cite{sycl2020, goli2020towards}.

SYCL interoperating with existing native objects is supported by either {\texttt host\_task} or {\texttt interop\_task} interfaces inside the command group scope.  
When using the {\texttt interop\_task} interface, the SYCL runtime system injects a task into the runtime DAG that will execute from the host, but ensures dependencies are satisfied 
on the device.
This allows code within a kernel scope to be written as though it were running directly at the low-level API on the host, but produces side-effects on the device, \textit{e.g.}, external
API or library function calls. 

There are several implementations of SYCL API available including ComputeCpp~\cite{computecpp} that currently supports the SYCL 1.2.1 specification,
DPC++ and hipSYCL which incorporate SYCL 2020 features, such as unified shared memory (USM), and triSYCL~\cite{gozillon2020trisycl} which provides SYCL supports for FPGAs.

\section{SYCL-based RNG implementations of NVIDIA and AMD GPUs in \onemkl}
\label{sec:support-curand}
\subsection{Technical aspects}
\label{ssec:technical}
The integration of third-party RNG backends within \onemkl{} depends primarily on compiler support for (a) SYCL 2020 interoperability and (b) generating the specific intermediate representation for a given architecture's source
code.
To enable RNG on NVIDIA and AMD GPUs, one requires SYCL compilers supporting parallel thread (PTX) and Radeon Open Compute (ROCm) execution instruction set architectures which are used in the CUDA and AMD programming environment, respectively.
The PTX backend support is available in Intel's open-source LLVM project, and the ROCm backend is supported by the
hipSYCL LLVM project.

\begin{figure}[!ht]
\begin{center}
\includegraphics[width=\columnwidth]{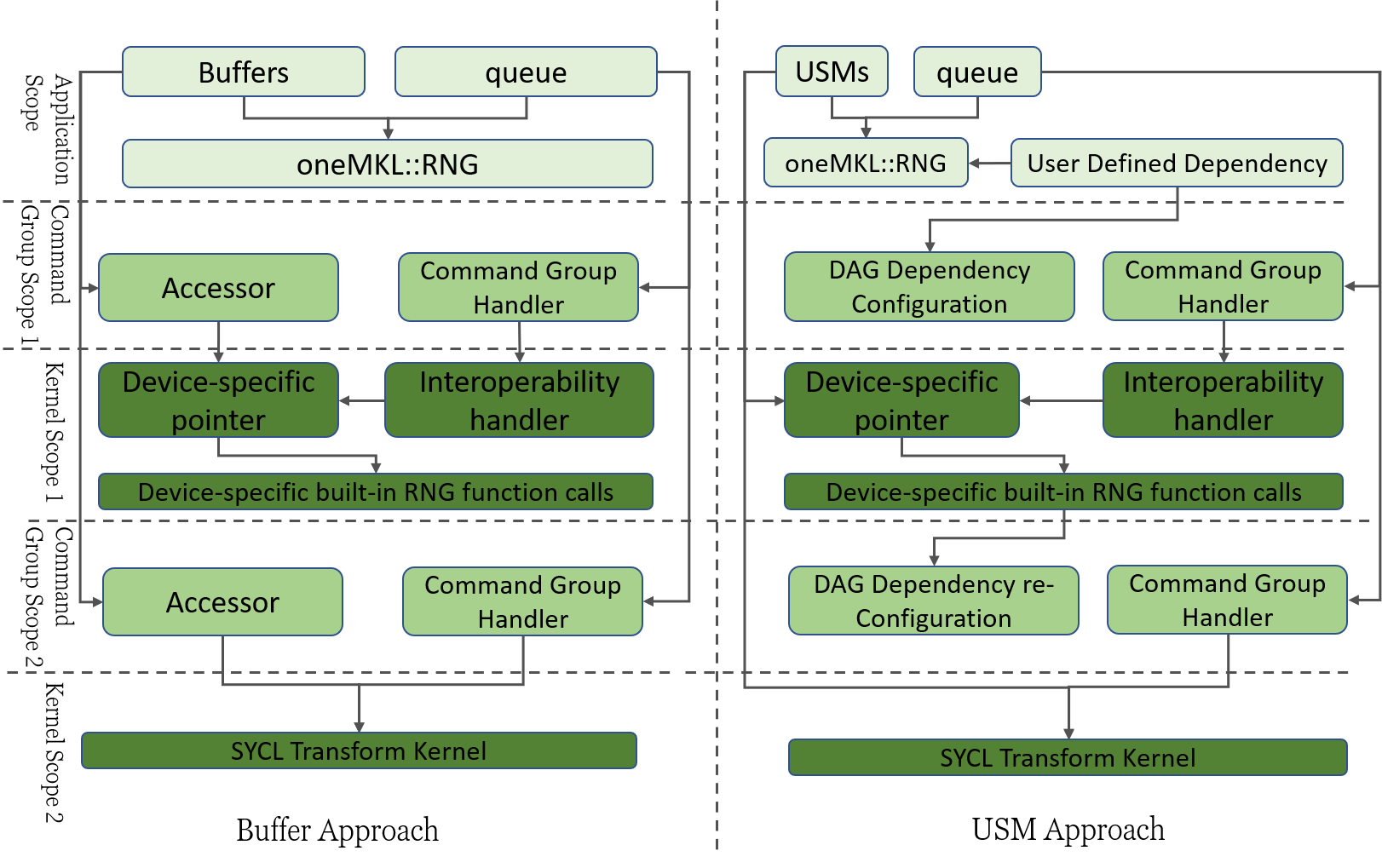}
\end{center}
\caption{Architectural view of device-specific RNG kernels integration in \onemkl{} for both \curand{} and \hiprand{} on different scopes in SYCL programming model using both buffer and USM approach.}
\label{fig:impl-arch}
\end{figure}

The \onemkl{} interface library uses both buffer and USM APIs for memory management.
Buffers are encapsulating objects which hide the details of pointer-based memory management.
They provide a simple yet powerful way for the SYCL runtime system to handle data dependencies between kernels, both on the host and device, when building the data-flow DAG.
The USM API gives a more traditional pointer-based approach, \textit{e.g.}, memory allocations performed with \texttt{malloc} and
\texttt{malloc\_device}, familiar to those accustomed to C++ and CUDA, \textit{e.g.}, \texttt{cudaMalloc}.
However, unlike buffers, the SYCL runtime system cannot generate the data dependency graph from USM alone, and so it is the user's responsibility to ensure dependencies are met.
The ability for SYCL to internally satisfy buffer-based data dependencies is beneficial in cases when
quick prototyping is to first order more important than optimizing.
Figure~\ref{fig:impl-arch} represents the architectural view of the \curand{} and \hiprand{} integration for each scope in the SYCL programming model for both buffer-based approach and USM-based approaches.

The \onemkl{} library currently contains implementations for Philox- and MRG-based generators for x86 and Intel GPU.
In \onemkl{}, each engine class comprises 36 high-level \texttt{generate} function templates -- 18 per buffer and USM API -- with template parameters to
specify a distribution and output types.
In addition to having the ability to specify distribution properties, \textit{e.g.}, mean, standard deviation for Gaussian
distributions, custom ranges on the generated numbers can also be specified.
This is in sharp contrast to the lower level interfaces provided by \curand{} or \hiprand{};
generation of random numbers is performed using functions with fixed types, and there is no concept of a ``range''; it is left to the user to post-process the
generated numbers.
For example, \texttt{curandGenerateNormal} will output a sequence of normally-distributed pseudorandom numbers in $[0, 1)$ and there
is no API functionality to transform the range.
As such, native \curand{} and \hiprand{} support generation of strictly positive-valued numbers.

Lastly, whereas \onemkl{} provides copy-constructors and constructors for setting seed initializer lists for multiple sequences, \curand{} and \hiprand{} do not.
The \onemkl{} library also supports inverse cumulative distribution function (ICDF) methods for pseudorandom number generation, while such methods are
available only for quasirandom number generators in the \curand{} and \hiprand{} API.

\subsection{Native \curand{} and \hiprand{} flow}
\label{ssec:curand-flow}
Generation of random numbers with \curand{} and \hiprand{} host APIs in native applications typically has the following workflow:
\begin{enumerate}
    \item the creation of a generator of a desired type;
    \item setting generator options, \textit{e.g.}, seed, offset, \textit{etc.};
    \item allocation of memory on the device using \texttt{\{cuda,hip\}Malloc};
    \item generation of the random numbers using a generation function, \textit{e.g.},\\
    \texttt{\{cu,hip\}randGenerate}; and
    \item clean up by calling the generator destructor\\ \texttt{\{cu,hip\}randDestroyGenerator} and \texttt{\{cuda,hip\}Free}.
\end{enumerate}
In addition, a user may wish to use the generated numbers on the host, in which case host memory must also be allocated and data transferred between devices.

\subsection{Implementation of \curand{} and \hiprand{} in \onemkl}
\label{ssec:impl-curand}
Our implementation of \curand{} and \hiprand{} within \onemkl{} follows closely the procedure outlined in Section~\ref{ssec:curand-flow}.
We also include additional \textit{range transformation} kernels for specifying the output sequence of random numbers, a feature not available in the \curand{} and \hiprand{} APIs.

Each generator class comprises a native \texttt{curandGenerator\_t} object.
Class constructors create the generator via a native  \texttt{curandCreateGenerator} API call and sets the seed for generation of the output sequence with
\texttt{curandSetPseudoRandomGeneratorSeed};
due to limitations of the \curand{} and \hiprand{} host API, our implementation does not support copy-construction or seed initializer lists.
Of the total 36 \texttt{generate} functions available in \onemkl{}, 20 are supported by
our \curand{} backend as the remaining 16 use ICDF methods (see
Section~\ref{ssec:technical}).
Each \texttt{generate} function in the \curand{} and \hiprand{} backends have the same signature as the corresponding x86 and Intel GPU function to facilitate ``pointer-to-implementation.''

The buffer and USM API \texttt{generate} function implementations are nearly identical;
access to the buffer pointer via a SYCL \texttt{accessor} is needed before retrieving the native CUDA memory.

\begin{lstlisting}[caption=Example code calling functions from the \curand{} library within a SYCL kernel using the buffer API., label=lst-kernel,float,floatplacement=ht]
virtual inline void generate(
  const oneapi::mkl::rng::uniform<float, uniform_method::standard>& distr,
  std::int64_t n, cl::sycl::buffer<float, 1>& r) override {
    queue_.submit([&](cl::sycl::handler& cgh) {
      auto acc = r.get_access<cl::sycl::access::mode::read_write>(cgh);
      cgh.codeplay_host_task([=](cl::sycl::interop_handle ih) {
      auto r_ptr = reinterpret_cast<float*>(
        ih.get_native_mem<cl::sycl::backend::cuda>(acc));
      curandStatus_t status;
      CURAND_CALL(curandGenerateUniform, status, engine_, r_ptr, n);
      cudaError_t err;
      CUDA_CALL(cudaDeviceSynchronize, err);
    });
  });
  range_transform_fp<float>(queue_, distr.a(), distr.b(), n, r);
}
\end{lstlisting}

\begin{lstlisting}[caption=Example code of transform function for \curand{} using the buffer API. The function can be used to transform the range of the generated numbers. Its dependencies are detected via the auto-generated runtime DAG graph from SYCL accessors.,  label=lst-kernel-transform,float,floatplacement=ht]
template <typename T>
static inline void range_transform_fp(cl::sycl::queue& queue, T a, T b,
                                      std::int64_t n,
                                      cl::sycl::buffer<T, 1>& r) {
  queue.submit([&](cl::sycl::handler& cgh) {
    auto acc =
      r.template get_access<cl::sycl::access::mode::read_write>(cgh);
    cgh.parallel_for(cl::sycl::range<1>(n), [=](cl::sycl::id<1> id) {
      acc[id] = acc[id] * (b - a) + a;
    });
  });
}
\end{lstlisting}

As shown in Figure~\ref{fig:impl-arch}, \curand{} and \hiprand{} backend integration in \onemkl{} requires two kernels.
The first kernel calls the corresponding \texttt{curandGenerate}, as per the distribution function template parameter type. Listing~\ref{lst-kernel} states the proposed kernel for \curand{} using the  buffer API. A second kernel is required to adjust the range of the generated numbers, altering the output sequence as required. Since this feature is not available in the \curand{} and \hiprand{} APIs, we have implemented this kernel directly in SYCL. Listing~\ref{lst-kernel-transform} gives an example of one such transformation kernel \curand{} using the buffer API.
In the command group scope, an accessor is required for the buffer API to track the kernel dependency and memory access within the kernel scope.
In this case, the graph dependency between the two kernels are automatically detected by a SYCL runtime system scheduling thread, tracking the data-flow based on the data access type, \textit{e.g}, \texttt{read}, \texttt{write}, \texttt{read\_write}.
The output accessor of the interoperability kernel has a \texttt{read\_write} access type and is passed as an input with \texttt{read\_write} access type to the transform kernel.
This forces the transformation kernel to depend on the SYCL interoperability kernel and hence the kernels will be scheduled for execution in this order.

The USM API does not require \texttt{accessor}s in the command group scope, but does take an additional argument for specifying dependent kernels for subsequent kernels using the data outputted.
The dependency is preserved by a direct injection of the \texttt{event} object returned by the command group handler to the existing dependency list.  

Inside the kernel scope for both buffer and USM APIs, calls to the \curand{} API are made from the host and, if using buffers, the \texttt{accessor} is then reinterpreted as native memory --
\textit{i.e.}, a raw pointer to be used for \curand{} and \hiprand{} API calls.
The random numbers are then generated by calling the appropriate \texttt{curandGenerate} as per the distribution function template parameter type.

The application's scope remains the same as the one proposed in the \onemkl{} SYCL RNG interface for both buffer and USM API, enabling users to seamlessly execute application on AMD and NVIDIA GPUs with no code modification whatever.

\section{Benchmark applications}
\label{sec:benchmark-applications}
Two benchmark applications were used for performance portability studies, and are detailed below.
The SYCL codes were compiled using the \texttt{sycl-nightly-20210330} tag of the Intel LLVM open-source DPC++ compiler for targeting CUDA devices and hipSYCL v0.9.0 for AMD GPUs. The applications' native counterparts were compiled with \texttt{nvcc} 10.2 and \texttt{hipcc} 4.0, respectively, for NVIDIA and AMD targets.
Calls to the high-resolution \texttt{std::chronos} clock were bootstrapped at different points of program execution to measure the execution time of different routines in the codes.

\subsection{Random number generation burner}
\label{ssec:benchmark-rngburn}
The first application was designed as an artificial benchmark to stress the
hardware used in the experiments by generating a sequence of pseudorandom numbers of a given batch size using a specified
API -- \textit{i.e.}, CUDA, HIP or SYCL -- and platform.
We use this simple test as the primary measure of our \onemkl{} RNG implementations.
Having a single application to benchmark all available platforms has a number of advantages, namely, ensuring
ease of consistency among the separate target platform APIs, \textit{e.g.}, all memory allocations, and data
transfers between host and devices are performed analogously for each API.

The workflow of this benchmark application can be outlined as follows:
\begin{enumerate}
    \item target platform, API and generator type are chosen at compile-time, specified by \texttt{ifdef} directives;
    \item target distribution from which to sample, number of iterations and cardinality of the output pseudorandom sequence are specified at runtime;
    for SYCL targets, buffer or USM API is also specified;
    \item host and device memory are allocated, and the generator is constructed and initialized; for SYCL targets, a
    distribution object is also created as per Step 2 above;
    \item pseudorandom output sequence is generated and its range is transformed; and
    \item the output sequence is copied from device memory to host memory.
\end{enumerate}

\subsection{FastCaloSim}
\label{ssec:benchmark-fcs}
Our second benchmark is a real-world application that aims to solve a real-world problem:
rapid production of sufficiently accurate high-energy collider physics simulations.
The parameterized calorimeter simulation software, FastCaloSim~\cite{Schaarschmidt2017}, was developed
by the ATLAS Experiment~\cite{Aad:1129811} for this reason.
The primary ATLAS detector comprises three sub-detectors;
from inner radii outward, a silicon-based inner tracking detector;
two types of calorimeter technologies consisting of liquid argon or scintillating tiles for measurements of traversing
particles' energies;
and at the periphery a muon spectrometer.
Among these three sub-detectors, the simulation of the calorimeters are the most CPU-intensive due to the complex
 -- \textit{i.e.} production of additional particles in particle-material interactions -- and stopping of highly energetic particles, predominantly in the liquid argon calorimeters.

The original FastCaloSim codes, written in standard C++, were ported to CUDA and Kokkos~\cite{10.3389/fdata.2021.665783}, and subsequently to SYCL;
the three ports were written to be as similar as possible in their kernels and program flow so as to permit  comparisons
between their execution and runtimes.
The SYCL port, largely inspired in its design by the CUDA version, permits execution on AMD, Intel and
NVIDIA hardware, whereas the CUDA port permits execution on NVIDIA GPUs exclusively.

We briefly describe the core functionality of FastCaloSim here;
for more details on the C++ codes and CUDA port, the reader is referred respectively to \cite{Schaarschmidt2017} and
\cite{10.3389/fdata.2021.665783}.
The detector geometry includes nearly 190,000 sensitive elements,
$\mathcal{O}(10)$ MB, each of which can record a
fraction of a traversing particle's energy.
Various parameterization inputs, $\mathcal{O}(1)$ GB, are used for different particles' energy and shower shapes,
derived from Geant4 simulations.
The detector geometry, about 20 MB of data, is loaded onto the GPU;
due to the large file size of the parameterization inputs, only those data required -- based on the particle type and kinematics -- are transferred during runtime.

The number of calorimeter \textit{hits} (\textit{i.e.} energy deposits in the sensitive elements) depends largely on the physics process being simulated.
For a given physics event, the number of secondary particles produced can range from one to $\mathcal{O}(10^4)$, depending on the incident parent particle type, energy
and location in the calorimeter.
Three uniformly-distributed pseudorandom numbers are required for each hit to sample
from the relevant energy distribution, with the minumum set to 200,000 --
approximately one per calorimeter cell.

We consider two different simulation scenarios in our performance measurements.
The first is an input sample of $10^3$ single-electron events, where each electron
carries a kinetic energy of 65~GeV and traverses a small angular region of the calorimeters.
An average number of hits from this sample is typically 4000--6500, leading to
12000--19500 random numbers per event.
Because only a single particle type is used within a limited region of the detector,
this scenario only requires a single energy and shower shape parameterization to be
loaded onto the GPU during runtime.
The second, more realistic, scenario uses an input of 500 top quark pair ($t\bar{t}$)
events.
In this simulation, the number of calorimeter hits is roughly 600--800 times greater
than the single-electron case, requiring
$\mathcal{O}(10^7)$ random numbers in total be generated during simulation.
Also, a range of secondary particles are produced with various energies that traverse a range of angular regions of the detector.
As such, $t\bar{t}$ simulations require data from 20--30 separate parameterizations that need to
be loaded to the GPU during runtime, and thus result in a significant increases in time-to-solution on both CPUs and GPUs.

\section{Performance evaluation}
\label{sec:perfport}

\subsection{Performance portability metrics}
There are numerous definitions of performance portability, \textit{e.g.}~\cite{zhu2007performance, edwards2014kokkos,mcintosh2014performance,larkin2016performance}.
In this paper, we adopt the definition from~\cite{pennycook2019implications}:
the performance portability $\mathcal{P}$ of an application
$a$ that solves a problem $p$ correctly on all platforms in a given set $H$ is
given by,
\begin{equation}
    \mathcal{P}(a, p; H) =
        \begin{cases}
            \frac{|H|}{\sum_{i \in H} \tfrac{1}{e_i(a, p)}} & \textrm{if } i \textrm{ is supported } \forall_i \in H \\
            0 & \textrm{otherwise}
        \end{cases},
\end{equation}
where $e_i(a, p)$ is the \textit{performance efficiency} of $a$ solving $p$ on $i \in H$. \\
We introduce an \textit{application efficiency} metric, the ratio between the time-to-solution ($TTS$) measured using our portable, vendor-agnostic (VA) solution to the native, vendor-specific (VS) performance:
\begin{equation}
    \text{VAVS} \equiv \frac{TTS_\text{portable}}{TTS_\text{native}}
\end{equation}
The VAVS metric is useful to identify if runtime overheads are introduced in portability layers which
otherwise do not exist in a native API optimized for a specific platform.

\subsection{Hardware specifications}
\label{ssec:testhardware}
We evaluate performance portability using a variety of AMD, Intel and NVIDIA platforms, ranging from consumer-grade to high-end hardware.
This large set of platforms can be subdivided into CPUs and GPUs, as well as the union of the two, and also helps determine the regime in which the use of GPUs is more efficient for solving a given problem, if one exists.

The Intel x86-based platform tested was a Core i7-10875, consisting of 8 physical CPU
cores and 16 threads, a base (maximum) clock frequency of 2.30 (5.10) GHz.
To benchmark native \onemkl{} GPU performance, we use the Intel(R) UHD Graphics 630, an
integrated GPU (iGPU) that shares the same silicon die as the host CPU described
previously.
This iGPU has 24 compute units (CU) and base (maximum) frequency of 350 (1200) MHz.
Through Intel's unified memory architecture (UMA), the iGPU has a theoretical maximum
memory of 24.98 GB -- \textit{i.e.}, the total available RAM on the host.
The main advantage of UMA is that it enables zero-copy buffer transfers -- no buffer
copy between the host and iGPU is required since physical memory is shared between
them.

We evaluated SYCL interoperability for AMD and NVIDIA GPUs using an MSI Radeon RX Vega 56 and NVIDIA A100.
The Radeon is hosted by an Intel Xeon Gold 5220 36-core processor with a base (maximum)
clock of 2.2 (3.9)~GHz.
An AMD CPU and NVIDIA GPU were evaluated using a DGX A100 node, comprising an AMD Rome
7742 64-core processor with a base (maximum) clock frequency of 2.25 (3.4)~GHz.
The A100 is NVIDIA's latest high-end GPU, with 6912 CUDA cores and peak FP32 (FP64) of
19.5 (9.7)~TF.
Note that 16 CPU cores and a single A100 of the DGX were used for these studies.
 
\subsection{Software specifications}
\label{ssec:testsoftware}
The software used for these studies can be found in Table~\ref{table:software}.
As our work is relevant only for Linux operating systems (OS), all test machines
run some flavor of Linux that supports the underlying hardware and software
required for our studies.
In this table, DPC++ refers to the Intel LLVM compiler nightly tag from March 3, 2021;
distinct builds of the compiler were used for x86 platforms and NVIDIA GPUs.
The HIP compiler and hipSYCL are based on Clang 12.0.0, and were installed from pre-compiled binaries available
from~\cite{hipsycl-rpm}.

Our implementations of SYCL-based \curand{} and \hiprand{} RNGs within \onemkl{}
were compiled into separate libraries for each platform using the respective compiler for
the targeted vendor.

\begin{table*}[t]
    \centering
    \begin{tabularx}{\columnwidth}{c||c|c|c|c}
 Platform &Driver Version &OS and Kernel &Compiler &RNG Library\\
 \hline
 \hline
  \multirow{2}{*}{AMD Rome 7742} &\multirow{2}{*}{-} &OpenSUSE 15.0 &GNU 8.2.0 &CLHEP 2.3.4.6 \\
   &&4.12 &DPC++ &\onemkl{} \\
 \hline
  \multirow{2}{*}{Intel Core i7-1080H} &\multirow{2}{*}{-} &Ubuntu 20.04 &GNU 8.4.0 &CLHEP 2.3.4.6\\
   &&5.8.18 &DPC++ &\onemkl{}\\
 \hline
  \multirow{2}{*}{Intel UHD Graphics} &\multirow{2}{*}{21.11.19310} &Ubuntu 20.04 &\multirow{2}{*}{DPC++} &\multirow{2}{*}{\onemkl{}}\\ && 5.8.18 & \\
 \hline
  \multirow{2}{*}{Radeon RX Vega 56} &\multirow{2}{*}{20.50} &CentOS 7 &HIP 4.0.0 &hipRAND 4.0.0\\
  &&3.10.0 &hipSYCL 0.9.0 &\onemkl{} \\
 \hline
  \multirow{2}{*}{NVIDIA A100} &\multirow{2}{*}{450.102.04} &OpenSUSE 15.0 &CUDA 10.2.89 &\curand{} 10.2.89\\
  &&4.12 &DPC++ &\onemkl{}\\
    \end{tabularx}
    \caption{Driver and software versions for each platform considered in these studies.}
    \label{table:software}
\end{table*}

\section{Results}
\label{sec:results}
The RNG burner application was run 100 iterations for each batch size for statistically meaningful measurements.
Each test shown in the following was performed with the Philox4x32x10 generator
to produce uniformly-distributed FP32 pseudorandom numbers in batches between
$1$--$10^8$, as per the requirements of our FastCaloSim benchmark application.
Unless otherwise specified, all measurements are of the total execution time, which includes generator construction, memory allocation, host-to-device data transfers, generation and post-processing (\textit{i.e.}, range transformations),
synchronisation and finally device-to-host data transfer times, as determined by
the high-resolution \texttt{std::chronos} clock.

\begin{figure}[!ht]
  \captionsetup[subfigure]{justification=centering}
\begin{center}
\subfloat[]{\includegraphics[width=\columnwidth]{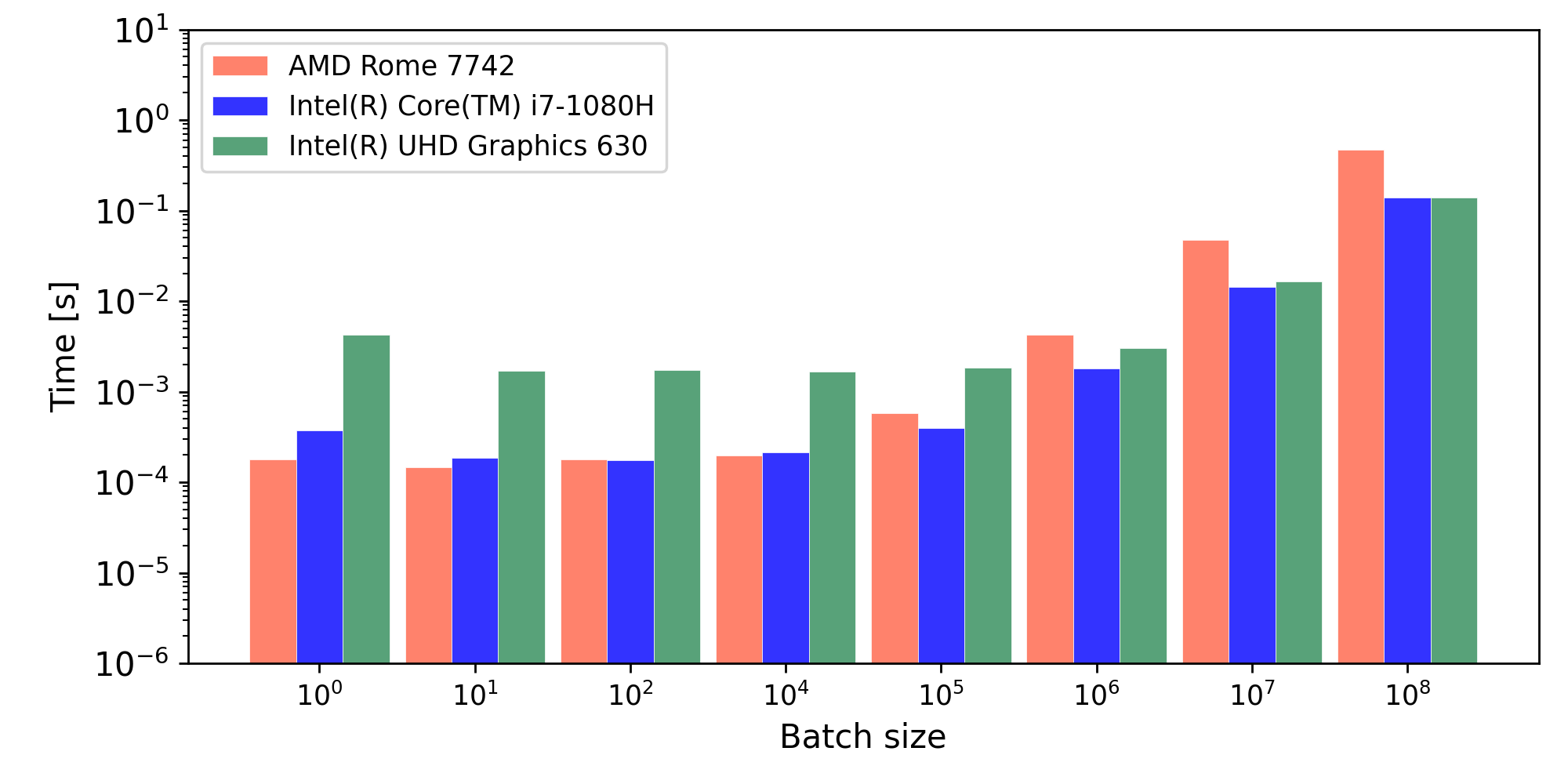}} \\
\subfloat[]{\includegraphics[width=\columnwidth]{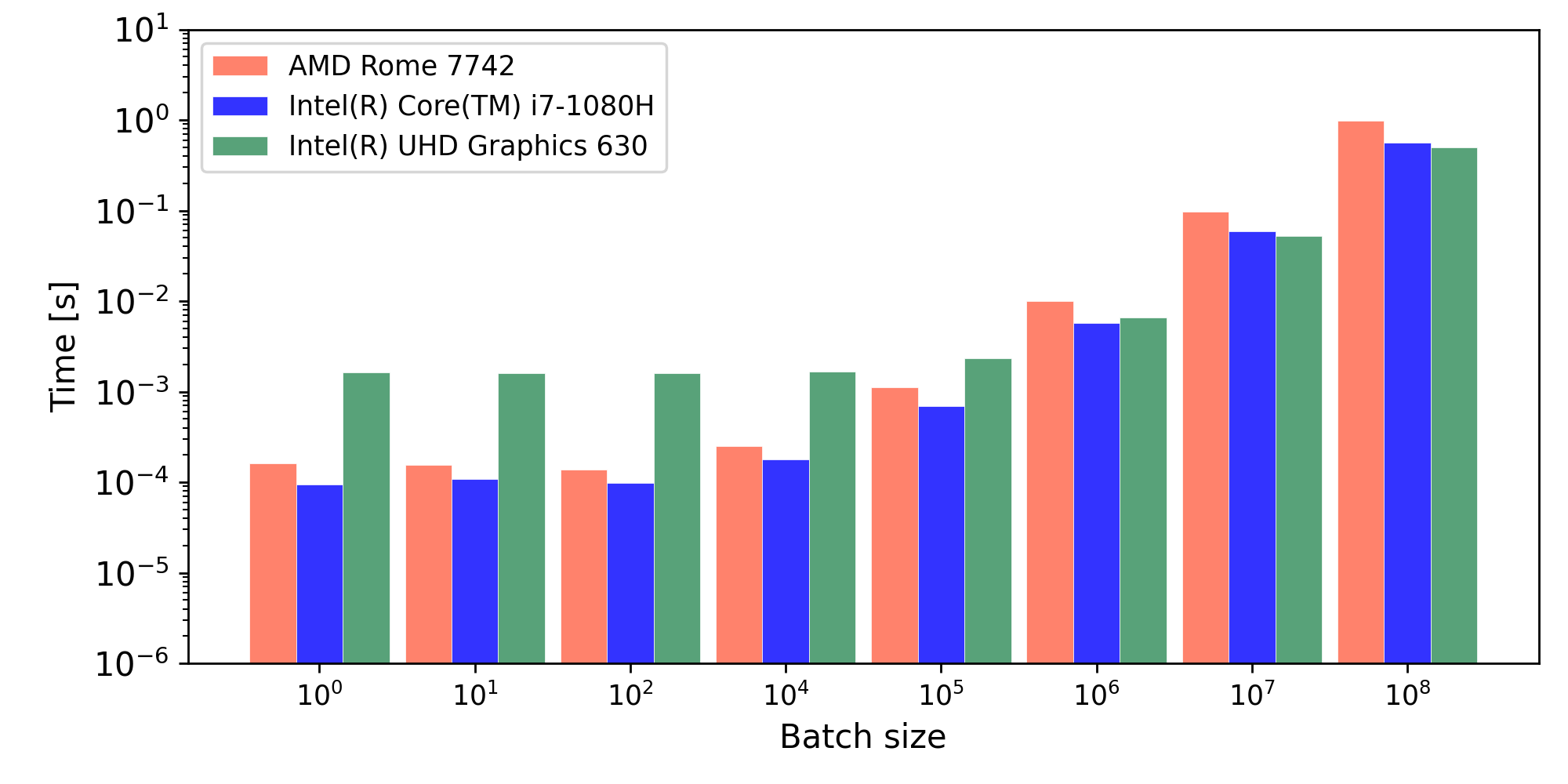}}
\end{center}
\caption{Results from the RNG burner test application using the buffer API (a) and USM API (b) for Philox4x32x10 generation of uniformly-distributed FP32 pseudorandom numbers.
}
\label{fig:tet-cpu}
\end{figure}

Shown in Figure~\ref{fig:tet-cpu} are plots of the total FP32 generation time for the two x86-based CPUs, as well the integrated GPU, using Philox-based generator for both buffer and USM APIs.
In general, little overhead is introduced when using the USM API versus buffers.
This is a promising result and, to the authors' knowledge, the first benchmark of the different APIs;
it is often more productive for developers to port existing codes to SYCL using USM as this approach is often
more familiar to C++ programmers who use dynamic memory allocations in their applications.

\begin{figure}[!ht]
  \captionsetup[subfigure]{justification=centering}
\begin{center}
\subfloat[]{\includegraphics[width=\columnwidth]{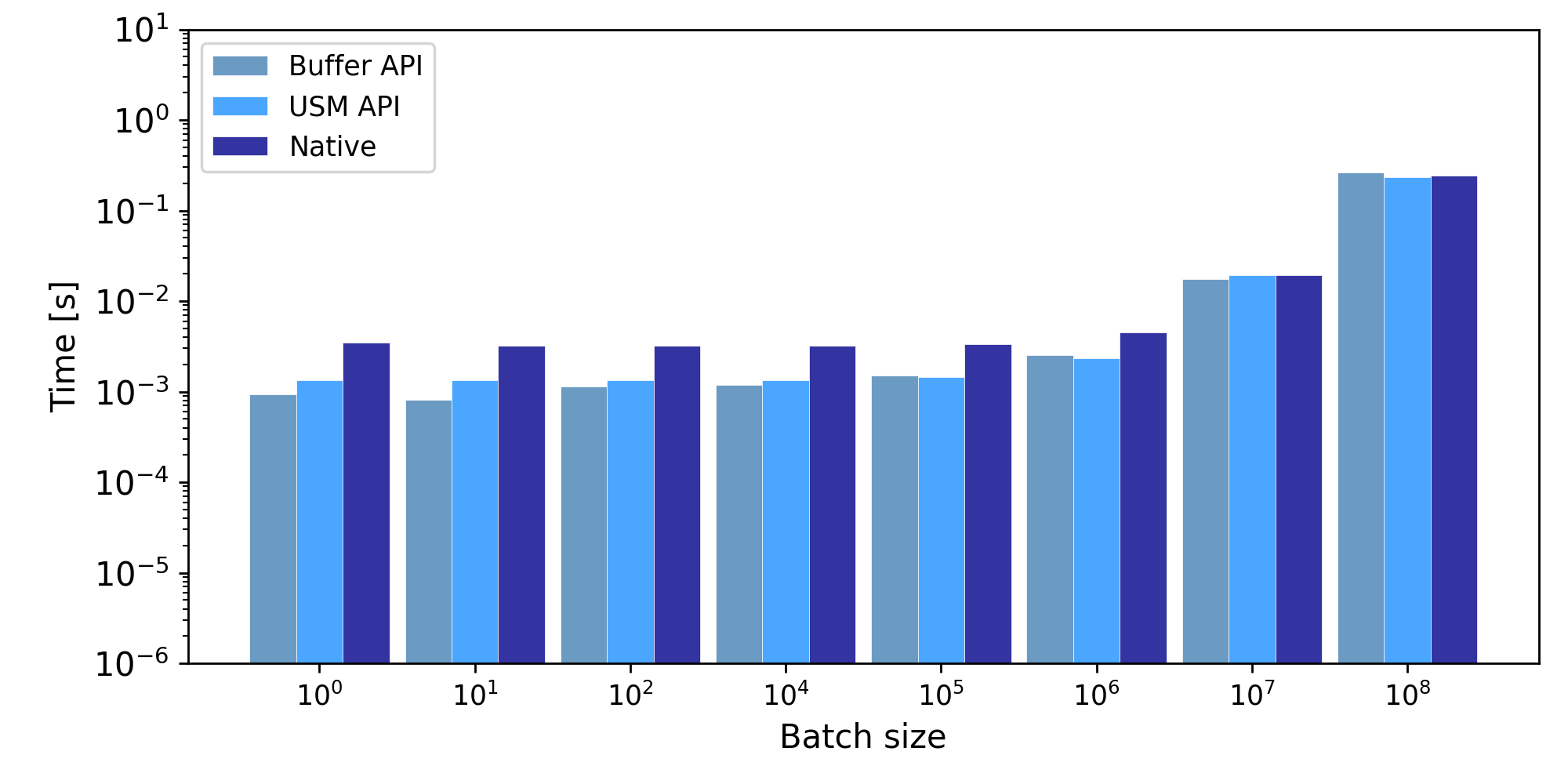}} \\
\subfloat[]{\includegraphics[width=\columnwidth]{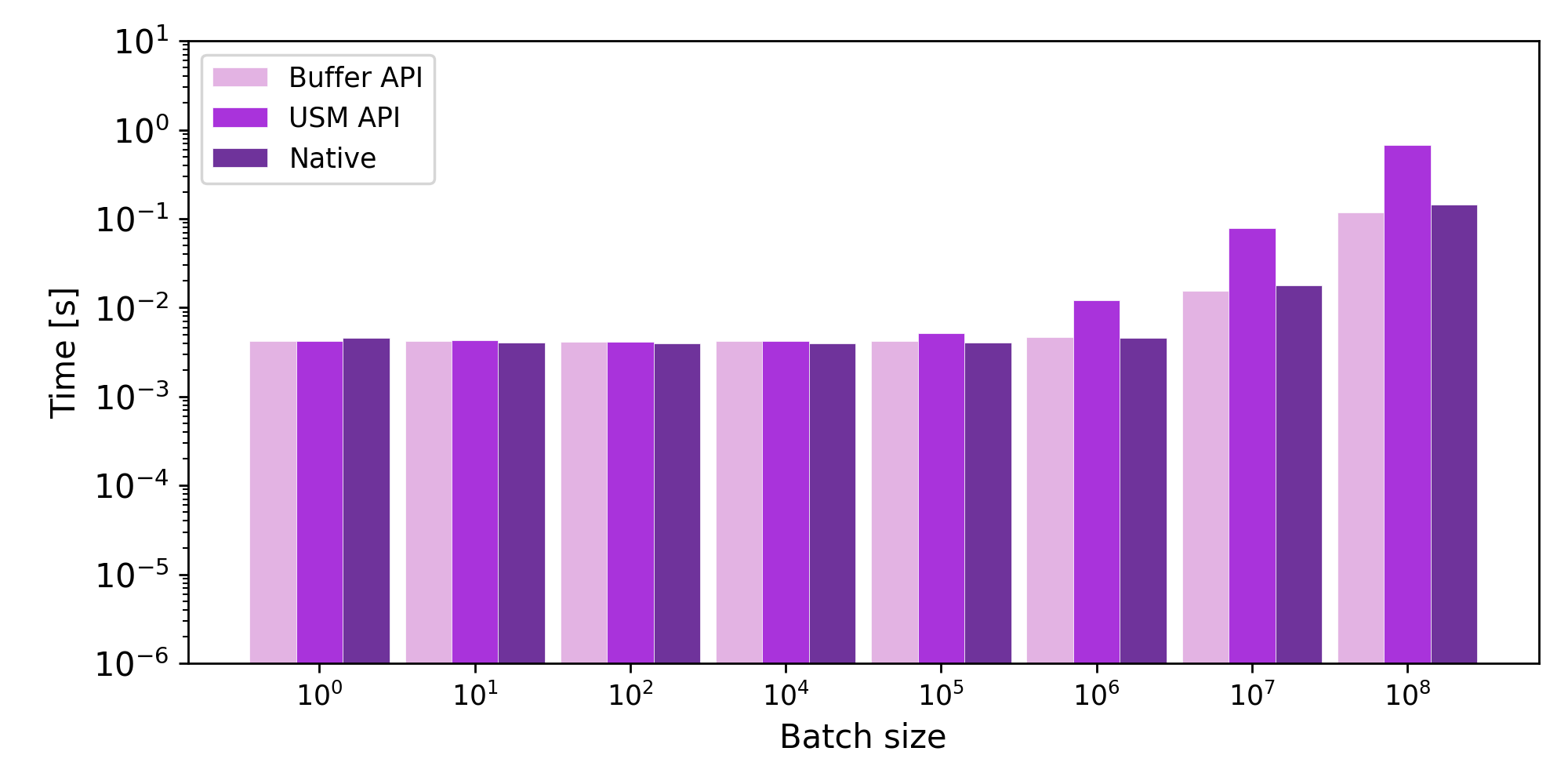}}
\end{center}
\caption{Comparisons of the RNG burner test application execution time between SYCL buffer and USM APIs, and their native counterparts running on the MSI Radeon RX Vega 56 (a) and NVIDIA A100 (b).
The Philox4x32x10 generator was used to produce uniformly-distributed FP32 pseudorandom numbers of different batch sizes.
}
\label{fig:tet-gpu}
\end{figure}

Figure~\ref{fig:tet-gpu} shows separately the RNG burner application results between the buffer and USM APIs, and their native
counterparts.
Again, we observe statistically equivalent performance using either buffers or USM, with
a slight overhead at large batch sizes DPC++ USM and the A100 GPU.
More importantly, however, is the level of performance achieved by our cross-platform
RNG implementation;
$TTS$ for both the \curand{} and \hiprand{} SYCL backend implementations
are on par with their native application.
%

One immediate point of discussion are the differences in $TTS$
between the Radeon \onemkl{}-based generator application and native
application:
the \onemkl{} version shows slightly better performance for small batch sizes.
This is understood as being a result of the optimizations within the \hiprand{} runtime system for its ROCm back-end.
Due to the data dependencies among the three kernels -- seeding, generation and post-processing -- in the test application,
call-backs are issued to signal task completion.
These call-backs introduce latencies into the application execution that are significant with respect to small-scale kernels.
The nearly callback-free \hiprand{} runtime system therefore offers higher task throughput.
As the batch sizes increase to $10^8$, the difference in $TTS$ becomes negligible.

\begin{figure}[!ht]
  \captionsetup[subfigure]{justification=centering}
\begin{center}
\subfloat[]{\includegraphics[width=\columnwidth]{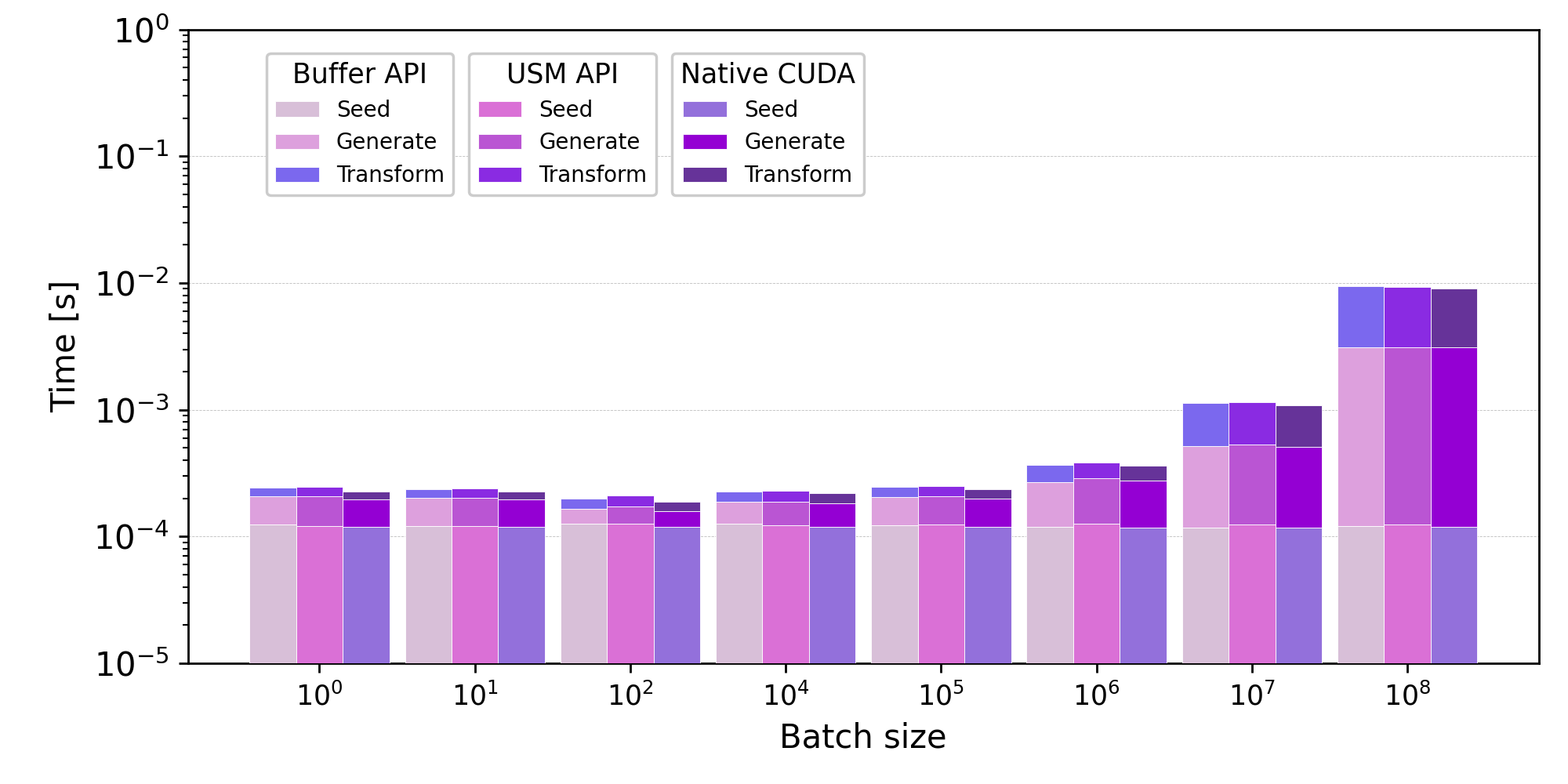}} \\
\subfloat[]{\includegraphics[width=\columnwidth]{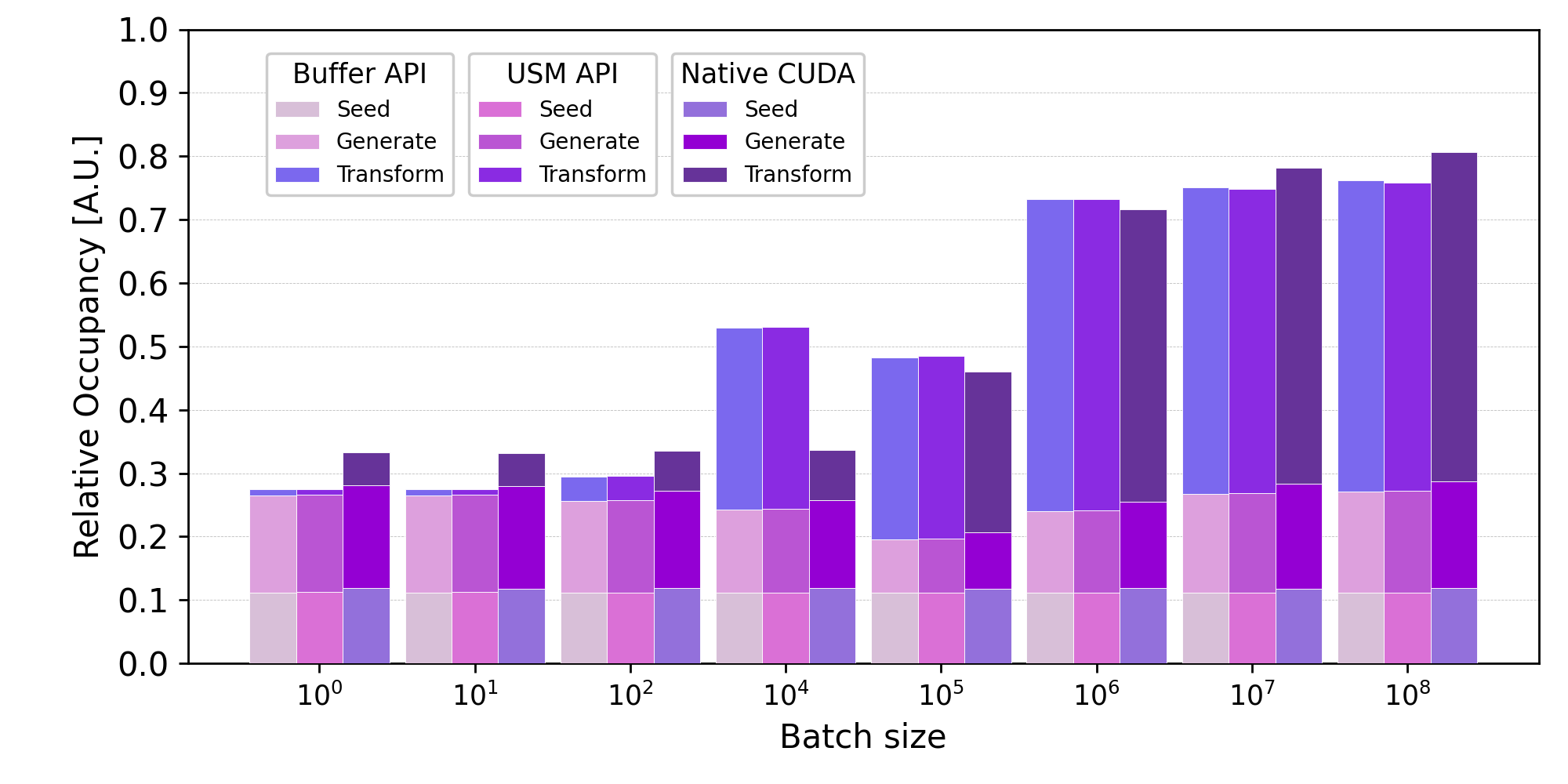}}
\end{center}
\caption{Per-kernel total execution time (a) and relative occupancy (b) executed on the NVIDIA A100 with the Philox4x32x10 generator producing uniformly-distributed pseudorandom sequences of various batch sizes.
}
\label{fig:ket}
\end{figure}

To further investigate this discrepancy, we separate each kernel's duration
for both the \onemkl{} and native \curand{} applications;
due to technical and software limitations, we were unable to profile the Radeon GPU in the
following way.
Three kernels in total are profiled: generator seeding, generation and our
transformation kernel that post-processes the output sequence to the defined range.
Figure~\ref{fig:ket} shows both the time of each kernel executed and relative occupancy in the RNG
burner application using data collected from NVIDIA Nsight Compute 2020.2.1.
Comparison between each kernel duration is statistically compatible over a series
of ten runs.
It can therefore conclude that the discrepancies in Figure~\ref{fig:tet-gpu} between the
Radeon \onemkl{} and native applications can be attributed to differences between the
applications themselves, and not fundamentally to the native library kernel executions.
Shown also in Figure~\ref{fig:ket}(b) are the relative occupancy of each kernel for the batch
sizes generated.
Both \curand{} kernels -- seeding and generation -- are in all cases statistically
equivalent between \onemkl{} and the native application.
It can be seen that, despite the nearly identical kernel duration, the buffer and USM API occupancies have a large increase between $10^2$ and $10^4$ in batch size compared to the native occupancy.
This is because when not explicitly specified, the SYCL runtime system optimizes the number of
required block size and threads-per-block, whereas in CUDA these values must be determined by the developer as per the hardware
specifications.
While in the native version the thread-per-block size is fixed to 256, the SYCL kernel runtime chose 1024 for the NVIDIA A100 GPU.
This resulted in the observed differences in kernel occupancy in the native application, as opposed to the SYCL codes for the transform kernel which handle such intricacies at the device level.

Table~\ref{table:pp-calc} reports the calculated performance portability of our \onemkl{} RNG backends using the VAVS metric introduced in Section~\ref{sec:perfport}. Note that VAVS values closer to unity are representative of greater performance, while
smaller values are indicative of poor performance.
The data used in calculating the various values of $\mathcal{P}$ are taken from Figure~\ref{fig:ket}.
\begin{table}[h]
    \centering
    \begin{tabularx}{27em}{c|c|c|c}
 $H$ & $\mathcal{P}$ buffer &$\mathcal{P}$ USM &$\mathcal{P}$ Mean (buffer+USM)\\
 \hline
 \hline
  \{Vega 56, A100\} & 1.070 & 0.393 & 0.575 \\
  \{Vega 56\} &0.974 &1.076 &1.022 \\
  \{A100\} &1.186 &0.240 &0.400
    \end{tabularx}
    \caption{Calculated performance portability using the VAVS metric.}
    \label{table:pp-calc}
\end{table}

As reported in Table~\ref{table:pp-calc}, the performance portability measure in a number of cases
is greater than unity.
This result is consistent with the performance improvement over the the native version observed in Figure~\ref{fig:tet-gpu} for the buffer API on both AMD and NVIDIA GPUs.
Although the interoperability kernel time is the same in both native and SYCL versions (see Figure~\ref{fig:ket}(a)), the buffer API leverages the SYCL runtime system DAG mechanism and hipSYCL optimizations, improving throughput relative to the native
application, particularly for small batch sizes.
On the other hand, the DPC++ runtime system scheduler does not perform the same with USM as it does when using buffers.
Therefore, the performance drop observed in the USM version in Figure~\ref{fig:tet-gpu} leads to a reduction in the performance portability metric by $\sim$\%40. This behaviour is not observed with hipSYCL. 

\begin{figure}[!ht]
  \captionsetup[subfigure]{justification=centering}
\begin{center}
\subfloat[]{\includegraphics[width=\columnwidth]{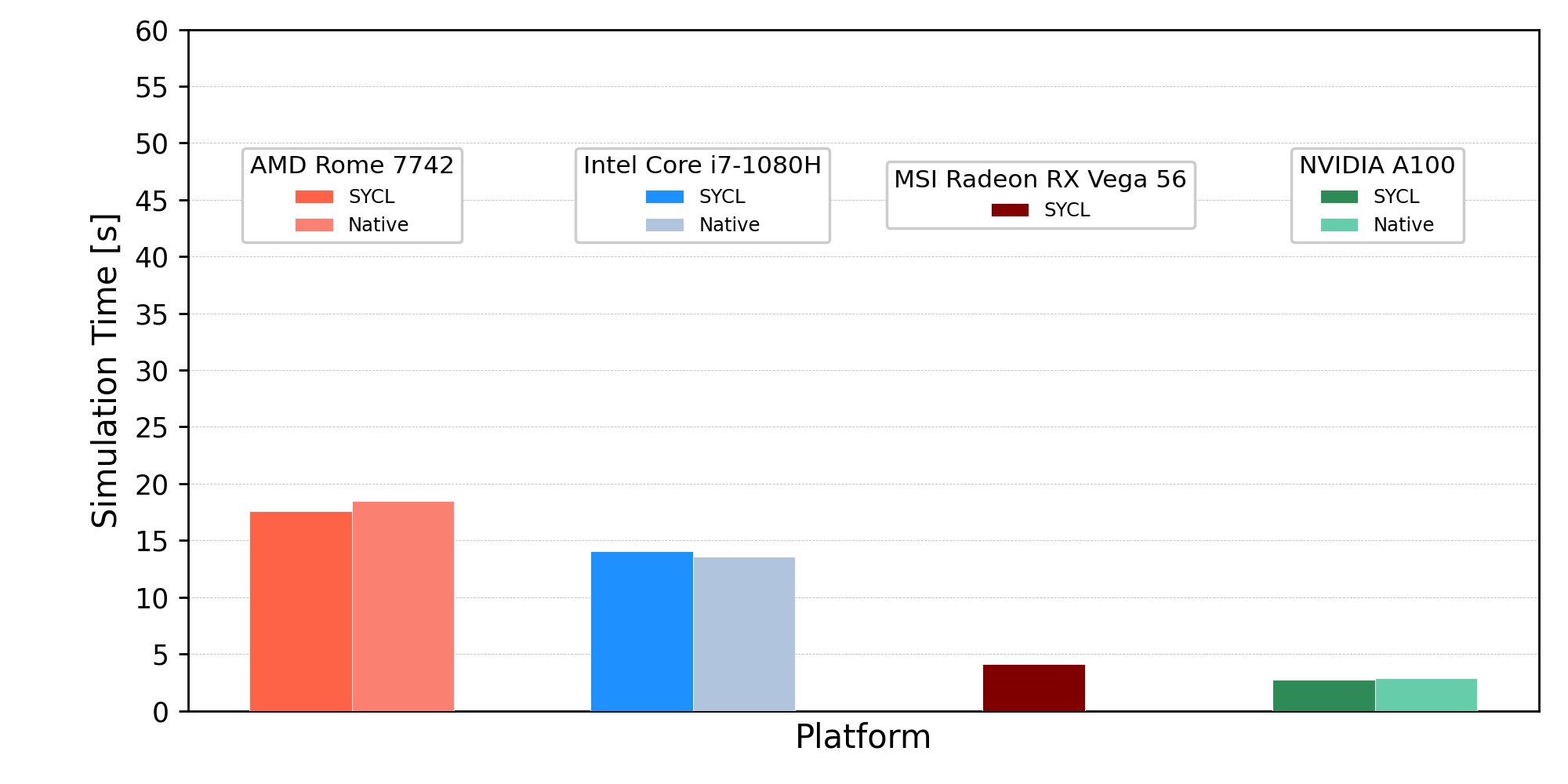}} \\
\subfloat[]{\includegraphics[width=\columnwidth]{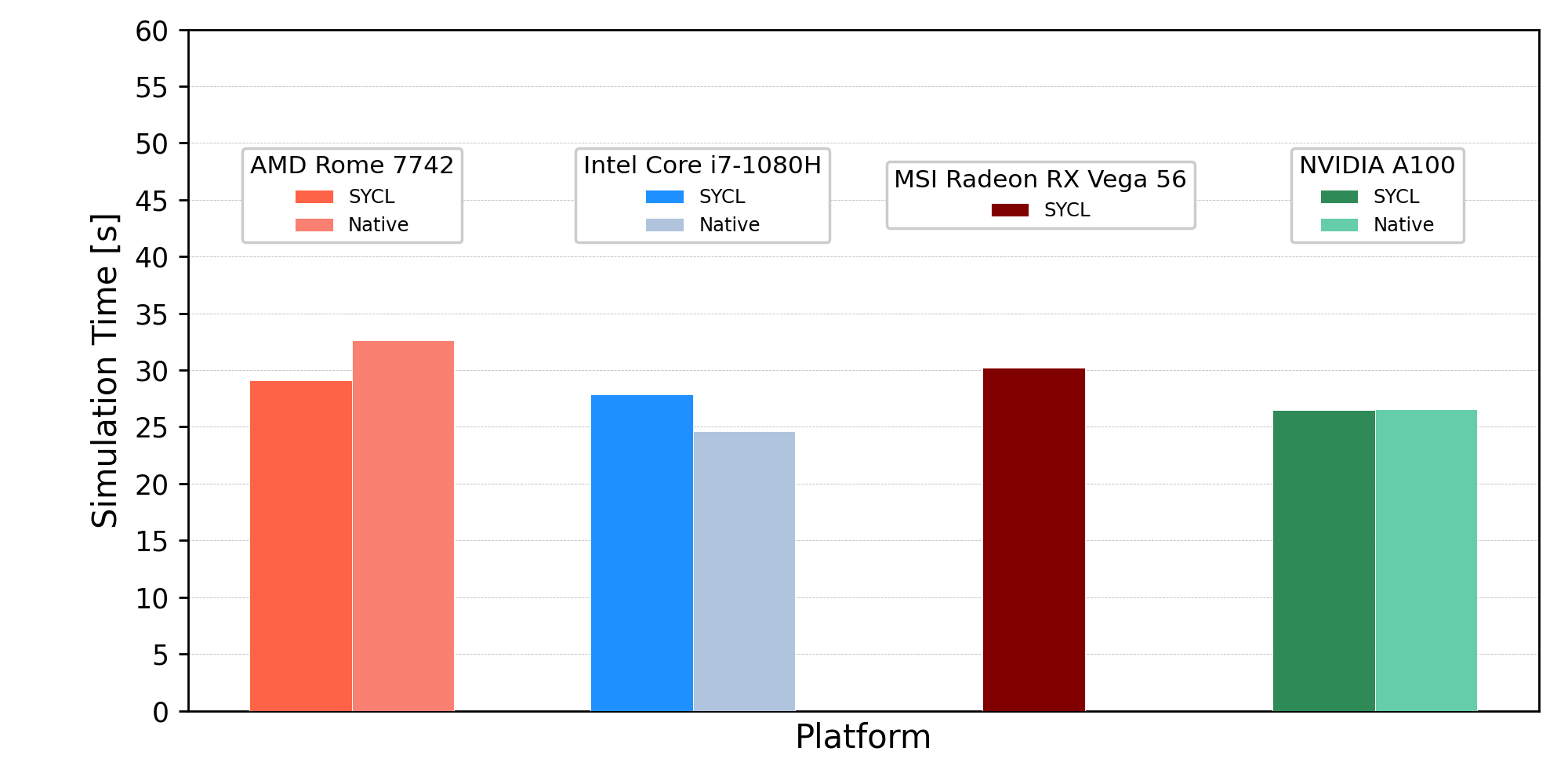}}
\end{center}
\caption{Total runtimes of FastCaloSim across a range of platforms simulating
single-electron events (a) and $t\bar{t}$ events (b).
}
\label{fig:fcs}
\end{figure}

As a demonstration of cross-platform performance portability in a real-world
application, we show in Figure~\ref{fig:fcs} the average runtime of the FastCaloSim code implementing the proposed SYCL RNG solution across four platforms.
Both SYCL and native implementations are shown for each platform, with the exception of the
Radeon GPU as no native HIP-based port exists.
Ten single-electron and $t\bar{t}$ simulations were run on each platform for reliability of measurements.
Where applicable, all measurements made in this study are consistent with those in~\cite{10.3389/fdata.2021.665783}.
The left plot in the figure pertains to the 10,000 single-electron events and the
right to the 500 $t\bar{t}$ events (see Section~\ref{ssec:benchmark-fcs}).

In the simpler scenario of single electrons, an approximately 80\% reduction in processing
time is required on the Vega or A100 GPUs compared to the CPUs considered.
However, the overall insufficient use of the full compute capability
of the GPUs in this application is made apparent in the more complex topology of $t\bar{t}$ events.
This inefficiency is due primarily to the initial strategy in porting FastCaloSim to GPUs;
while maximum intra-event parallelism -- \textit{i.e.} parallel processing of individual hits within a given event -- is met,
inter-event parallelism is not implemented in this version of the codes.
Future work on the FastCaloSim ports includes event batching to better utilize GPU compute but is beyond the scope of this
paper.
While the contribution of RNG to the overall runtime of FastCaloSim is small, to investigate SYCL as a portability solution for these codes nevertheless required a SYCL RNG to do so.
With \curand{} and \hiprand{} support added to \onemkl{}, we can run this prototype application on all major vendors'
platforms with no code modifications whatever, and with comparable performance to native
codes.

\section{Conclusions and future work}
\label{sec:conclusion}

In this paper, we detailed our implementations of \curand{} and \hiprand{} backends into \onemkl{}, and studied their cross-platform performance portability in two SYCL-based applications using major high performance computing hardware, including x86-based CPUs from AMD and Intel, and AMD, NVIDIA and Intel GPUs.
We have shown that utilizing SYCL interoperability enables performance portability of highly-optimized platform-dependent libraries across different hardware architectures.
The performance evaluation of our RNG codes carried out in this paper demonstrates little overhead when exploiting vendor-optimized native libraries through interoperability methods. Moreover, utilizing different SYCL compilers, DPC++ and hipSYCL, enabled the portability of the same SYCL code across different architectures with no modification in the application, highlighting the cross-platform portability of applications written in open-standard APIs across different hardware vendors.

The applicability of the proposed solution has been evaluated in a parameterized calorimeter
simulation software, FastCaloSim, a real-world application consisting of
thousands of lines of code and containing custom kernels in different languages and vendor-dependent libraries.
The interfaces provided by \onemkl{} enabled the seamless integration of SYCL RNGs into FastCaloSim with no code modification across the evaluated platforms.
The SYCL 2020 interoperability functionality enabled custom kernels and vendor-dependent library integration to be abstracted out from the application, improving the maintainability of the application and reducing the source lines of code.
The application yields comparable performance with the native approach on different architectures.
Whereas the ISO C++ version of FastCaloSim had two separate codebases for x86 architectures and NVIDIA GPUs, the work presented
here has enabled event processing on a variety of major vendor hardware from a single SYCL entry point.
Hence, the SYCL RNG based integration facilitates the code maintainability by reducing the FastCaloSim code size without introducing any significant performance overhead.

While we have demonstrated that SYCL interoperability leads to reusability of existing optimized vendor-dependent libraries and enables cross-platform portability, devices without vendor libraries cannot be supported.
For example, no RNG kernels exist yet for ARM Mali devices.
One possible solution would be to provide pure SYCL kernel implementations for common RNG engines.
The kernel could then be compiled for any device for which a SYCL-supported compiler exists.
Moreover, in scientific applications and workflows where reproducibility is essential,
kernels written entirely in the SYCL programming model can offer improved reliability across architectures and platforms.
Although the portability of such an RNG kernel would be guaranteed, performance remains challenging and likely would
necessitate mechanisms such as tuning of kernels for different architectures.

In future work, we plan to: (1) implement a number of mathematical kernels directly within SYCL; (2) benchmark their performance with respect to their respective vendor-dependent counterparts; and (3) extend the analysis to include productivity
(\textit{e.g.}, ease of integration into existing applications) and reproducibility.
In particular, extending the performance, portability and productivity  that also includes reproducibility in an objective way would  general scientific applications and workflows aiming for architecture and platform independence.





%
%
%
\bibliographystyle{splncs04}
\bibliography{ref}
%




\end{document}